\documentclass[aps,prb, amsmath,amssymb, superscriptaddress, twocolumn, showpacs]{revtex4-1}
\usepackage[english]{babel}
\usepackage{graphicx}
\usepackage{amsmath}
\usepackage{amssymb}
\usepackage[outdir=./]{epstopdf}
\usepackage{amsfonts}
\usepackage{bm}
\usepackage{mathtools}
\usepackage{times}
\usepackage{hyperref}
\usepackage{color}
\newcommand{\Neel}{N\'{e}el }

\newcommand{\beq}[1]{\begin{equation} #1 \end{equation}}
\newcommand{\bsplit}[1]{\begin{equation} \begin{split} #1 \end{split} \end{equation}}

\newcommand{\astcycl}{\mathrlap{\kern0.085em{\circlearrowright}}\ast}
\newcommand{\taucycl}{\mathrlap{\kern0.42em{\bullet}}\circlearrowright}
\long\def\/*#1*/{}

\newcommand{\+}{\dagger}
\def\<{\left\langle}
\def\>{\right\rangle}
\def\up{\uparrow}
\def\dn{\downarrow}

\def\P2d2h{\mathrm{P}_\mathrm{2d2h}}

\def\hH{\hat{H}}

\def\hd{\hat{d}}

\def\hc{\hat{c}}
\def\hn{\hat{n}}
\def\Snn{S_\mathrm{NN}}
\def\tmax{t_\mathrm{max}}

\newcommand{\vk}{\textbf{k}} \newcommand{\vK}{\textbf{K}}

\begin{document}

\title{Effects of frustration on the nonequilibrium dynamics of photo-excited lattice systems}
\author{Nikolaj Bittner}
\email{nikolaj.bittner@unifr.ch}
\affiliation{Department of Physics, University of Fribourg, 1700 Fribourg, Switzerland}
\author{Denis Gole\v{z}}
\affiliation{Flatiron Institute, Simons Foundation, 162 Fifth Avenue, New York, NY 10010, USA}
\author{Martin Eckstein}
\affiliation{Department of Physics, University of Erlangen-N\"urnberg, 91058 Erlangen, Germany}
\author{Philipp Werner}
\email{philipp.werner@unifr.ch}
\affiliation{Department of Physics, University of Fribourg, 1700 Fribourg, Switzerland}

\begin{abstract}
We theoretically investigate the effects of the lattice geometry on the nonequilibrium dynamics of photo-excited carriers in a half-filled two-dimensional Hubbard model. Using a nonequilibrium generalization of the dynamical cluster approximation, we compare the relaxation dynamics in lattices which interpolate between the triangular lattice and square lattice configuration and thus reveal the role of the geometric frustration in these strongly correlated nonequilibrium systems. In particular, we show that the cooling effect resulting from the disordering of the spin background is less effective in the triangular case because of the frustration. This manifests itself in a longer relaxation time of the photo-doped population, as measured by the time-resolved photo-emission signal, and a higher effective temperature of the photo-doped carriers in the non-thermal steady state after the intra-Hubbard-band thermalization.
\end{abstract}


\maketitle

\section{Introduction}

Geometric frustration can cause unusual effects and the appearance of exotic phases in strongly correlated materials.\cite{sahebsara2008,yoshioka2009,shirakawa2017,ye2019} This is the result of an inability of the system to minimize simultaneously all the interactions. A good illustration is an electron system with antiferromagnetic interactions on a two-dimensional triangular lattice. Placing two electrons on a triangle with anti-parallel spin orientations makes it impossible for the third one to form two antiferromagnetic bonds.
Lattice systems with geometric frustration often exhibit a large manifold of highly degenerate states,  and may therefore be highly susceptible to external perturbations. 

A potentially rich playground for studying geometrically frustrated systems are cold atoms in optical lattices, where all parameters can be adjusted independently. Recent advances in cooling procedures have enabled simulations of the two-dimensional (2D) Hubbard model and the observation of antiferromagnetic order,\cite{mazurenko2017} the structure of the chemically doped states,\cite{hilker2017} as well as anomalous transport~\cite{brown2017} at elevated temperatures. This progress may soon enable the study of exotic phases of matter, like quantum spin liquid~(QSL) states, that exhibit a variety of new features associated with their topological character.\cite{ balents2010} 
Moreover, recently several solids have been proposed to host QSL states, including organic materials,\cite{shimizu2003,kanoda2006,itou2007,yamashita2010,yamashita2011,isono2014} quantum kagome lattices\cite{han2012,norman2016colloquium,gomilvsek2019kondo} and other triangular systems like 1T-TaS$_2$.\cite{klanjvsek2017high,he2018} While the low-temperature properties of frustrated materials have been investigated for decades, much less is known about the effect of frustration on their nonequilibrium dynamics.\cite{ishikawa2014,kawakami2018} In view of the nontrivial underlying physics, it is interesting to identify the fingerprints of geometrical frustration in nonequilibrium probes.

An established approach for studying the properties of a strongly correlated electron system out of equilibrium is the nonequilibrium implementation~\cite{aoki2014_rev} of dynamical mean field theory (DMFT).~\cite{georges1996, metzner1989} This method, however, describes non-local correlations and interactions only at the mean field level, which is not sufficient to capture the effects of geometrical frustration. Extensions of DMFT, such as extended DMFT (EDMFT)~\cite{sun2002, ayral2013, werner2016b} or the combination of GW and EDMFT,\cite{biermann2003,werner2016a} were successfully used in nonequilibrium settings,~\cite{golez2015,golez2017} and in particular allowed to study the dynamical screening associated with nonlocal interactions.~\cite{golez2015,bittner2018}
While some of these methods capture long-range correlations, they cannot accurately describe the effects of strong short-range correlations,
which turn out to be important for the dynamics of frustrated systems. For the purpose of this work,  extensions of the DMFT approach which explicitly treat the short-range correlations within a small cluster are therefore more suitable.~\cite{hettler1998, maier2005} 
These cluster-based extensions have been extensively used in the equilibrum context and have already provided useful insights into the properties of the Hubbard model on a triangular lattice.~\cite{lee2008, chen2010,dang2015}
Implementations of cluster-based DMFT methods for out-of-equilibrium problems are however still rare and nonequilibrium studies of nonlocal correlation effects represent a frontier in this research field.~\cite{tsuji2014, eckstein2016, hermann2016, bittner2020} 

In the present work, we reveal possible fingerprints of geometric frustration in photo-doped Mott insulating systems. 
The photo-excitation of electrons across a large Mott gap leads to the creation of long-lived doublon (holon) type charge carriers in the upper~(lower) Hubbard band. Here, we investigate how their dynamics and quasi-steady-states are 
influenced by frustration. Specifically, we explore this question within 
Mott insulating 2D Hubbard models at half-filling, using a nonequilibrium generalization of the dynamical cluster approximation (DCA).~\cite{hettler1998, maier2005} We consider systems without long-range magnetic order, but treat the short-range spin correlations within the clusters. On non-frustrated lattices, the doublons and holons can efficiently dissipate their kinetic energy by reducing the short-range spin correlations that are present even above the \Neel temperature. Geometric frustration leads to a less effective spin cooling in the triangular case, compared to a square lattice geometry. We demonstrate this effect by analyzing  the time-dependent  spin correlation functions and the photoemission spectrum. Slow relaxation times of doublons and holons and their hot energy distributions in the non-thermal steady-states are nonequilibrium fingerprints of geometric frustration.

The paper is organized as follows. Section~\ref{sec:ModMed} describes the model and observables. Section~\ref{sec:Results} presents the results of our study for electronic systems with geometric frustration in and out of equilibrium, while Section~\ref{sec:Summary} summarizes the main findings. 

\section{Model and Method}
\label{sec:ModMed}

\begin{figure}[t]
\centering%
\includegraphics[width=1.\columnwidth, draft=false]{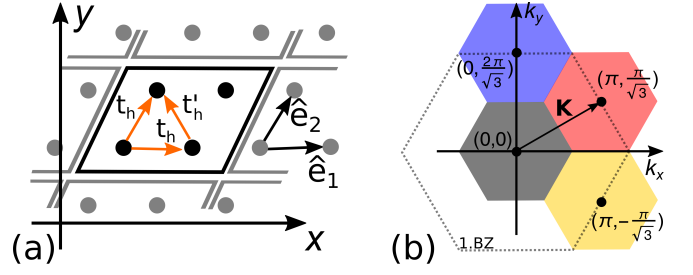}
\caption{ (a) Illustration of the triangular lattice in real space and of the direction-dependent electron hoppings. 
A real-space cluster with $N_c=2\times2$ sites is marked by the black line. 
(b) Corresponding reciprocal space representation with $N_c$ cells. The coarse graining around the momentum points $\vK$ in the first Brillouin zone (1. BZ) is indicated by the colored areas.
}
\label{fig:TriMesh}
\end{figure}

We consider a half-filled Hubbard model on a triangular lattice with anisotropic hopping parameters, which is described by the time-dependent Hamiltonian
\begin{align}
\label{eq:H}
	\hH(t)=&-\sum_{\<i,j\>\sigma}(t_{i,j}(t) \hc_{i\sigma}^\+\hc_{j\sigma}+\text{H.c.}) - \mu \sum_{i} \hn_i \nonumber\\
	& + U\sum_i \hn_{i\up}\hn_{i\dn}.
\end{align}
Here $\hc^{(\+)}_{i,\sigma}$ annihilates (creates) an electron on lattice site $i$  with spin $\sigma$ and $\<i,j\>$ represent pairs of nearest-neighbor sites. The density operator is $\hn_{i}=\hn_{i\up}+\hn_{i\dn}$ with $\hn_{i,\sigma}=\hc_{i,\sigma}^\+\hc_{i,\sigma}$, $U$ is the on-site interaction parameter, and $\mu$ the chemical potential. The hopping is direction dependent and takes the form (see Fig.~\ref{fig:TriMesh} (a)):
\begin{eqnarray}
	t_{i,j}(t)=\left\{
		\begin{array}{ll}
			t_h(t) & \text{ along } \hat e_1, \hat e_2\\
			t_h^\prime & \text{ along } \hat e_2 - \hat e_1
		\end{array}	
	\right.
\end{eqnarray}
where $t_h^\prime$ varies from the unfrustrated limit $t_h^\prime = 0$ (distorted square lattice) to the fully frustrated triangular case with $t_h^\prime=t_h$. The bare dispersion $\epsilon_\vk$ for the system with direction-dependent hopping parameters takes the form
\bsplit{
	\epsilon_\vk =&-2\left[ t_h \cos(k_x) +t_h\cos(\tfrac12 k_x+\tfrac{\sqrt{3}}{2}k_y)\right.\\
	 &\left.+t_h^\prime\cos(\tfrac12k_x-\tfrac{\sqrt{3}}{2}k_y) \right].
}
We note that the bandwidth $W$ of the noninteracting system depends on the degree of the geometrical frustration, namely $W=8t_h$ for $t_h^\prime=0$ and $W=9t_h$ for $t_h^\prime=t_h$. 

In order to study the relaxation dynamics of photo-doped doublons and holons in Mott insulating systems described by Eq.~\eqref{eq:H}, we employ the nonequilibrium generalization of the dynamical cluster approximation (DCA) in combination with a non-crossing approximation (NCA) impurity solver.~\cite{keiter1971,eckstein2010,bittner2020} 
Within this formalism, the cluster reference system (without bath) is represented by the cluster Hamiltonian
\bsplit{
	H^c(t)=&-\sum_{\<i,j\>\sigma}\left[t^{c}_{i,j}(t) \hd_{i\sigma}^\+\hd_{j\sigma}+\text{H.c.}\right] -\mu\sum_{i} \hn_{i} \\
	&	+U\sum_{i}  \hn_{i\up}\hn_{i\dn},
\label{eq:Hc}
}
with $\hd_{i\sigma}^{(\+)}$ denoting the annihilation (creation) operators on the cluster. 
The cluster hopping parameters $t_{i,j}^c$ ($t_{i,j}^c=6t_{i,j}/\pi^2$) are renormalized 
by the averaging over momentum patches in the DCA method, which also imposes translation invariance on the cluster. 
All the sites of this cluster are hybridized with a self-consistently determined bath. 

In this work, we use a $2\times2$ cluster, which corresponds to \mbox{$N_c=4$} patches in  reciprocal space (see Fig.~\ref{fig:TriMesh}) around the momentum points $\vK=\{(0 , 0), (\pi, \pi/\sqrt{3}), (\pi , -\pi/\sqrt{3}), (0, 2\pi/\sqrt{3})\}$.~\cite{imai2002,lee2008,dang2015}

The photo-doping of the large-gap Mott insulating state is generated by a time-dependent modulation of the hopping parameter along the $\hat{e}_1$ and $\hat{e}_2$ directions (see Fig.~\ref{fig:TriMesh}):
\begin{equation}
t_h(t) = t_h^0 +\Delta t_h e^{-(t-t_0)^2/\tau^2}\sin(\Omega_p (t-t_0)),
\label{eq_t_h}
\end{equation}
with amplitude $\Delta t_h$. 
The hopping modulation has a Gaussian envelope with a maximum at time $t_0$ and a full width at half maximum $\tau$, and the frequency $\Omega_p$ is chosen according to the gap size and the bandwidth. We measure the energy in units of the hopping $t_h^0$ of the unfrustrated system and time in units of $\hbar/t_h^0$. This excitation protocol is chosen because the  gauge-invariant formulation of DCA with electromagnetic fields has not yet been worked out. 
The details of the excitation protocol should not have qualitative effects on the discussed results:
The  effect of the excitation protocol is to generate a broad distribution of non-equilibrium carriers by an almost impulsive perturbation, while the main focus of this study is on the relaxation of carriers after the excitation.

\begin{figure*}[!htb]
\centering
\includegraphics[width=0.79\textwidth, draft=false]{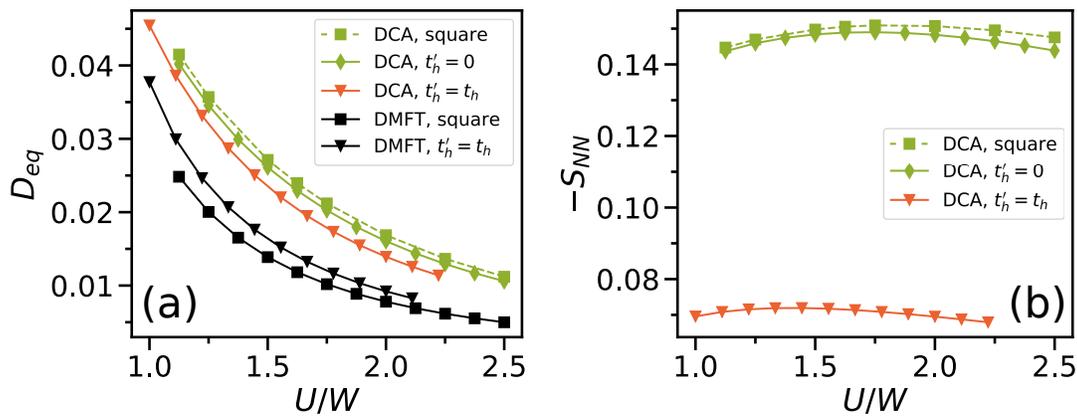}
\caption{(a) Double occupancy calculated at $T=0.1$ using DCA for different lattice geometries: the tilted square lattice ($t_h^\prime=0$, solid green line with diamonds), the square lattice (dashed green line with squares), and the triangular lattice ($t_h^\prime=t_h$, solid red line with triangles). For comparison, we also show results from DMFT calculations at the same temperature for the square (black line with squares) and triangular lattice (black line with triangles). The horizontal axis is normalized by the noninteracting bandwidth $W$. (b) Corresponding nearest neighbor spin-spin correlation function calculated using DCA with the same color coding as in panel (a).
}
\label{fig:EqDocc}
\end{figure*}

\subsection{Observables}
\label{sec_obs}
In this section, we define the observables that will be used to analyze the properties of these 2D lattice systems in and out of equilibrium. Whereas the single-particle observables are equivalent on the lattice and on the cluster, we measure the two-particle observables only on the cluster to avoid the computationally challenging solution of Bethe-Salpeter equations. The cluster results, 
e.g.~for 
the spin correlations, may not be identical to the lattice 
observables, but we expect them to reproduce the qualitative trends.~\cite{maier2005}

\subsubsection{Double occupation}
The double occupancy (or doublon density) $D(t)$ is defined as
\beq{
	D(t) = \frac{1}{N_c} \sum_{i=1}^{N_c} \<\hat{n}_{i\up} (t) \hat{n}_{i\dn}(t)\>  
\label{eq:docc}
}
with $i$ labeling the cluster sites.  
Within the DCA formalism, it is convenient to measure $D(t)$ in momentum space, as discussed in Ref.~\onlinecite{bittner2020}.  

\subsubsection{Spin-spin correlation function}

To measure the short-range spin correlations we calculate the nearest neighbor (NN) spin-spin correlation function on the $2\times2$ cluster. For the relevant cases of the fully frustrated triangular system ($t_h^\prime=t_h$) and the tilted square  lattice ($t_h^\prime=0$), we define the spin-spin correlation function as 
\beq{
	\Snn (t)=\frac{1}{N_{\delta}}\sum_{\<i,j\>}\<\hat{S}_i^z (t)\hat{S}_j^z (t)\>
\label{eq:SzSz}
}
with $\hat{S}_i^z$ denoting the spin operator in the $z$-direction at site $i$ and 
$N_{\delta}$ the number of pairs of neighboring sites connected by $t_h$, see also Ref.~\onlinecite{eckstein2016}. 
Taking into account the periodicity of the cluster, the geometrically frustrated system 
has $N_{\delta}=12$, while the tilted square lattice 
has $N_{\delta}=8$.

\subsubsection{Spectral functions}

To calculate the equilibrium spectral functions, we perform the Fourier transformation of the retarded component of the local Green's function:
\beq{
	A(\omega) = -\frac{1}{\pi} \mathrm{Im} \int_{0}^{\tmax} dt e^{-i\omega t} G^R(t)\label{eq:A}
}
with the Fourier time window $\tmax=10$ and $G^R=1/N_c\sum_\vK G^R_\vK$. 

We analyze the relaxation dynamics of the photo-doped population by computing the 
nonequilibrium photoemission spectrum~\cite{freericks09}
\begin{align}
\label{eq:spectrum}
	I(\omega, t_p)=&-\mathrm{Im}\int d\bar{t} d\bar{t'} e^{-i\omega(\bar{t}-\bar{t'})} G^< (t_p+\bar{t}, t_p+\bar{t'}) \nonumber \\
	&\times S(\bar{t}) S(\bar{t'}), 
\end{align}
where $S(t)\propto \exp[-(t^2/(2\Delta t_\mathrm{probe}^2)]$
is the envelope of the probe pulse of length $\Delta t_\mathrm{probe}=1.5$. 

\section{Results}
\label{sec:Results}
\subsection{Equilibrium properties}

In this section we discuss equilibrium properties of the half-filled 2D Hubbard models with different lattice geometries. The temperature of the system is set to $T=0.1$, unless otherwise specified,
and we focus on the paramagnetic Mott phase.

\subsubsection{Double occupancy}
\label{sSec:SSloc}

The $U$ dependence of the double occupancy $D_\text{eq}$ (Eq.~\eqref{eq:docc}) in equilibrium is plotted in Fig.~\ref{fig:EqDocc} (a) for a square lattice (green dashed line with squares), tilted square lattice (triangular lattice with $t_h^\prime=0$, green solid line with diamonds), and  triangular lattice with $t_h^\prime=t_h$ (red solid line with triangles). To take into account the different bandwidths $W$ of the electronic systems, we normalize the horizontal axis by $W$. 
The temperature $T=0.1$ is above the critical endpoint of the Mott transition, which occurs around $U/W\sim1$ for the triangular lattice~\cite{yoshioka2009,dang2015,shirakawa2017} and $U/W\sim0.5$  for the square lattice.~\cite{dang2015, eckstein2016} On all lattices, the double occupancies $D_\text{eq}$  are decreasing with increasing interaction $U$, however, there are quantitative differences between the lattices.

First, we compare the results for the square lattice with those for the tilted square lattice (triangular lattice with \mbox{$t_h^\prime=0$}). 
These two cases actually correspond to the same lattice Hamiltonian. The slight discrepancy in the results originates from averaging over different patches in the Brillouin zone (square-shaped patches centered at $\vK=\{(0 , 0), (\pi, 0), (0, \pi), (\pi, \pi)\}$ versus hexagonal patches centered at $\vK=\{(0 , 0), (\pi, \pi/\sqrt{3}), (\pi , -\pi/\sqrt{3}), (0, 2\pi/\sqrt{3})\}$), and therefore quantifies a DCA course graining error. In the fully frustrated triangular system with $t_h^\prime=t_h$ 
the double occupancy is suppressed compared to the square lattice case. This can be explained by the effect of the geometrical frustration, which suppresses antiferromagnetic spin correlations and thus increases the Pauli blocking of virtual hopping processes which contribute to $D_\text{eq}$. This effect will be further confirmed by analyzing the spin-spin correlation functions in the following subsection. 

It is also interesting to compare the DCA double occupancies within the corresponding single-site DMFT results, plotted by black lines in Fig.~\ref{fig:EqDocc}(a). We see that in the latter case the double occupancy is reduced. This can be explained by the fact that in contrast to DMFT, the DCA simulation captures non-local antiferromagnetic spin correlations, which reduce the Pauli blocking effect. Hence, nonlocal correlations have also a significant effect on local observables, such as the double occupation. We further notice that the effect of $t'_h$ on $D_\text{eq}$ is not correctly captured by single-site DMFT, which only takes into account the effect of $t'_h$ on the density of states, but misses the effect of frustration.

\subsubsection{Short range spin correlations}

The frustration picture is confirmed by analyzing the nearest-neighbor spin-spin correlation, see Eq.~\eqref{eq:SzSz}. The equilibrium results obtained using the DCA approximation are plotted in Fig.~\ref{fig:EqDocc}(b) as a function of $U/W$ for the square lattice, tilted square lattice ($t_h^\prime=0$), and triangular lattice ($t_h^\prime=t_h$). In the single-site DMFT limit the spin-spin correlation function vanishes in the paramagnetic phase, since this method does not capture short-range correlations.

Again, we see that the spin-spin correlation functions are very similar on the square lattice and the tilted square lattice, and the small numerical differences originate from the different patch geometries. However, in the triangular lattice case ($t_h^\prime=t_h$)  the spin-spin correlations in the $z$ direction are suppressed by about a factor of two due to the effect of geometrical frustration in the presence of antiferromagnetic nearest-neighbor correlations.

\subsubsection{Spectral function}

\begin{figure}[t]
\centering
\includegraphics[width=1.0\columnwidth, draft=false]{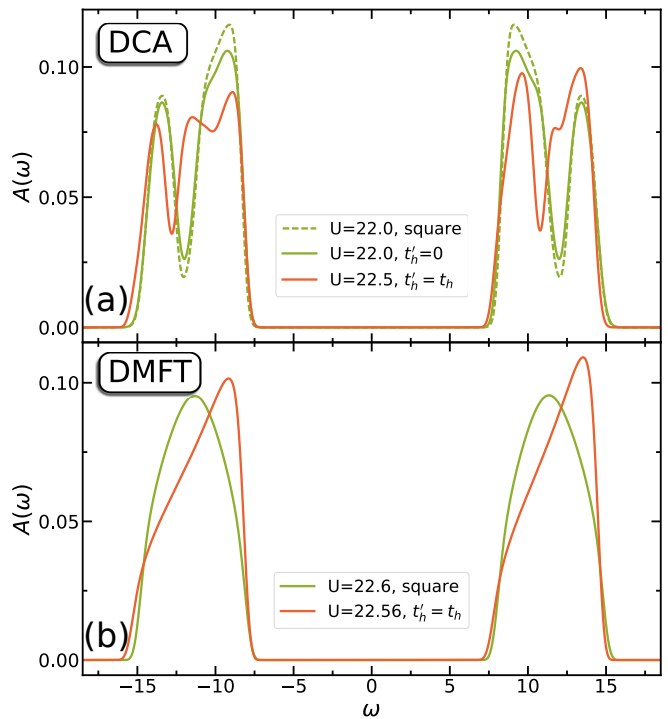}
\caption{Equilibrium spectral functions for different lattice geometries obtained using (a) the DCA and (b) the DMFT approximation. The interactions strengths $U$ and the hopping parameters $t_h$ have been adjusted to roughly match the gap size and the band width. (a) Spectral function for a tilted square lattice with $t_h^\prime=0$, $U=22$ (solid green line) and a triangular lattice~(solid red line) with renormalized parameters~($t_h^\prime=t_h\to0.8t_h^0$, $U=22.5$). For comparison, we also plot the spectral function for a square lattice with $U=22$ (dashed green line). (b) DMFT spectral function for the square lattice~(green solid line) with the renormalized  parameters~($t_h\to 0.87t_h^0$, $U=22.6$), and for the triangular lattice~(red solid line) with renormalized parameters~($t_h^\prime=t_h\to 0.76t_h^0$, $U=22.56$).
}
\label{fig:EqAw}
\end{figure}

In Fig.~\ref{fig:EqAw} (a), we plot the equilibrium spectral functions (see Eq.~\eqref{eq:A}) obtained using the DCA approximation for several lattice configurations: the square lattice (green dashed line), the tilted square lattice ($t_h^\prime=0$, green solid line), and the triangular lattice ($t_h^\prime=t_h$, red solid line). In order to eliminate effects associated with carrier recombination after the photo-excitation (discussed in Sec.~\ref{sec_nonequilibrium}), we choose the value of the interaction strength $U$ such that the corresponding Mott gap is almost twice as large as the bandwidth. 
Comparing $A(\omega)$ for the square and the tilted square lattice we see that the results are in a good agreement: (i) both calculations yield almost the same value of the Mott gap ($\Delta_\mathrm{Mott}\approx 15$), and (ii) the symmetric Hubbard bands~(with width $W\approx 8.2$) exhibit a similar two-peak structure. However, the details differ between the two spectra due to the averaging over different patches.

Including $t_h^\prime=t_h$ leads to a smaller value of the Mott gap and to asymmetric Hubbard bands, due to the lack of particle-hole symmetry in the triangular lattice. 
This results in different bandwidths of the lower Hubbard band ($W_\mathrm{LHB}\approx 8.6$) and upper  Hubbard band ($W_\mathrm{UHB}\approx 7.9$). 
For a meaningful comparison between the models, we adjust the interactions $U$ 
in the models with $t_h^\prime>0$ in such a way that the spectral functions roughly reproduce the gap size
of the model with $t_h^\prime=0$. 
Additionally, for a given degree of frustration $t_h^\prime>0$ the hopping parameter $t_h$ is renormalized with respect to its counterpart for the square lattice ($t_h\to\alpha t_h^0$ with a scaling factor $\alpha$), in order to roughly match the bandwidths of the models. 
Due to the asymmetry of the Hubbard bands in the triangular case, 
the resulting LHB is thus 
slightly broader than its counterpart for the square lattice, while the width of the upper Hubbard band is reduced by a corresponding amount. As an illustration, we plot in Fig.~\ref{fig:EqAw} the spectral function of the Hubbard model in the triangular case with the renormalized hopping parameters $t_h^\prime=t_h\to 0.8 t_h^0$ (red solid line). 

\begin{figure}[t]
\centering
\includegraphics[width=1.0\columnwidth, draft=false]{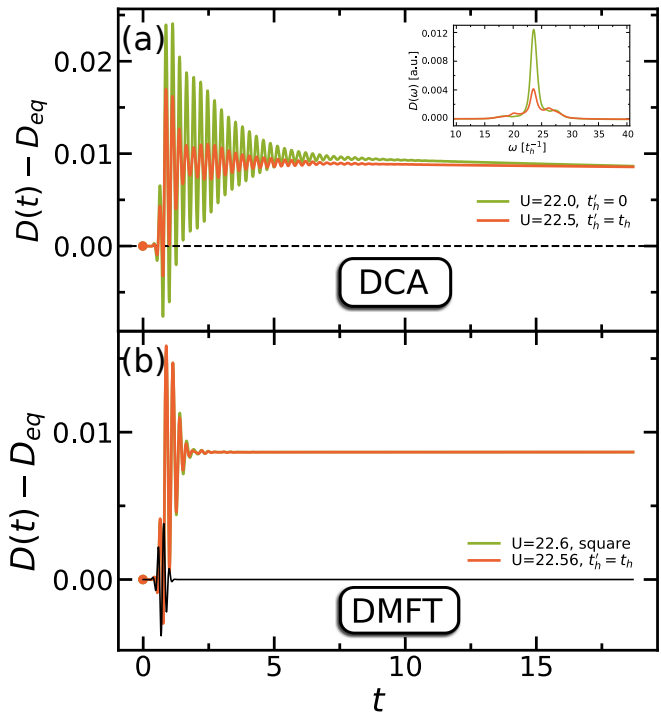}
\caption{Pulse-induced change in the double occupation. The parameters of the Mott-Hubbard systems are the same as in the equilibrium calculations (see Fig.~\ref{fig:EqAw}). (a) The DCA evolution of 
the double occupancy after a short excitation for the tilted square lattice~(green line) and the triangular lattice~(red line). The inset shows the Fourier transform of the data directly after pumping (with a smooth background subtracted) to illustrate the oscillations with a frequency $\omega\approx U$. (b) The DMFT evolution for the double occupancy on the square lattice~(green line) and on a
triangular lattice~(red line). The excitation strength is choosen such that the same photo-doping is achieved for the longest propagation time, namely $\Delta t_h\approx0.4$ for the tilted square lattice and $\Delta t_h\approx0.47$ for the triangular lattice in the DCA case. The excitation strengths in the DMFT case are $\Delta t_h\approx0.55$ and $\Delta t_h\approx0.45$ for the square and triangular lattices, respectively. The black line in (b) represents the pulse ($t_h(t)-t_h^0$) 
in arbitrary units.
}
\label{fig:NonEqdocc}
\end{figure}
 
We complement the DCA discussion by analyzing the single-site DMFT results, see Fig.~\ref{fig:EqAw}(b). Again, for a meaningful comparison between different models, we adjust the interactions $U$ and hopping parameters $t_h$ in order to reproduce the gaps and bandwidths of the spectral functions obtained using DCA. While  
the asymmetry~(symmetry) between the UHB and the LHB for the triangular~(square) lattice is captured by the DMFT spectrum, the fine structure of the Hubbard bands is different. The fine structure is influenced by short-range correlations, 
whereas the splitting of the Hubbard bands into subbands is likely to be over-emphasized in the 4-site DCA, as a result of the coarse discretization of the self-energy in momentum space. 

In the following nonequilibrium analysis, we will fix the parameters for each lattice as discussed in this section and only vary the parameters of the hopping modulation (Eq.~\eqref{eq_t_h}).

\subsection{Nonequilibrium properties}
\label{sec_nonequilibrium}

A hopping modulation of the form (\ref{eq_t_h}) with appropriate frequency creates charge excitations across the Mott gap. For the parameters of the photo-excitation we use a pulse with $\Omega_p=29$ centered at $t_0\approx0.73$, with a full width at half maximum $\tau=0.2$ (the pulse lasts up to $t\approx 1.8$, see Fig.~\ref{fig:NonEqdocc}). 
Here, we are interested in the relaxation dynamics of the photo-induced doublons and holons, which results from electron-electron scattering and interactions with the spin background. The dynamics is analyzed by means of the time-dependent observables introduced in Sec.~\ref{sec_obs}. 

\begin{figure}[b]
\centering
\includegraphics[width=1.\columnwidth, draft=false]{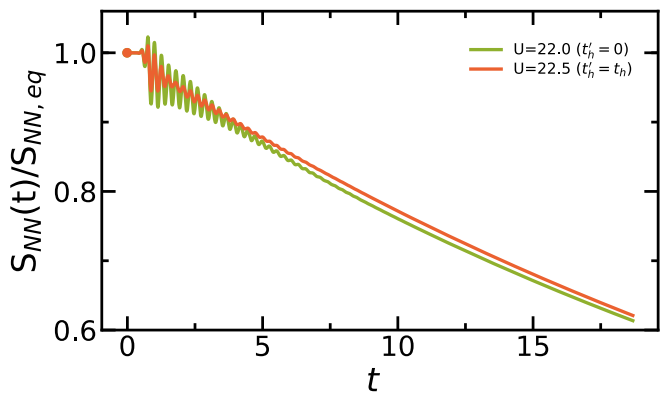}
\caption{Nearest-neighbor spin-spin correlation function (normalized by its equilibrium value) for a tilted square lattice~(solid green line), and a triangular lattice~(solid red line). The excitation protocols are the same as in Fig.~\ref{fig:NonEqdocc} (a). }
\label{fig:noneqSzSz}
\end{figure}

\begin{figure*}[ht]
\centering
\includegraphics[width=0.9\textwidth, draft=false]{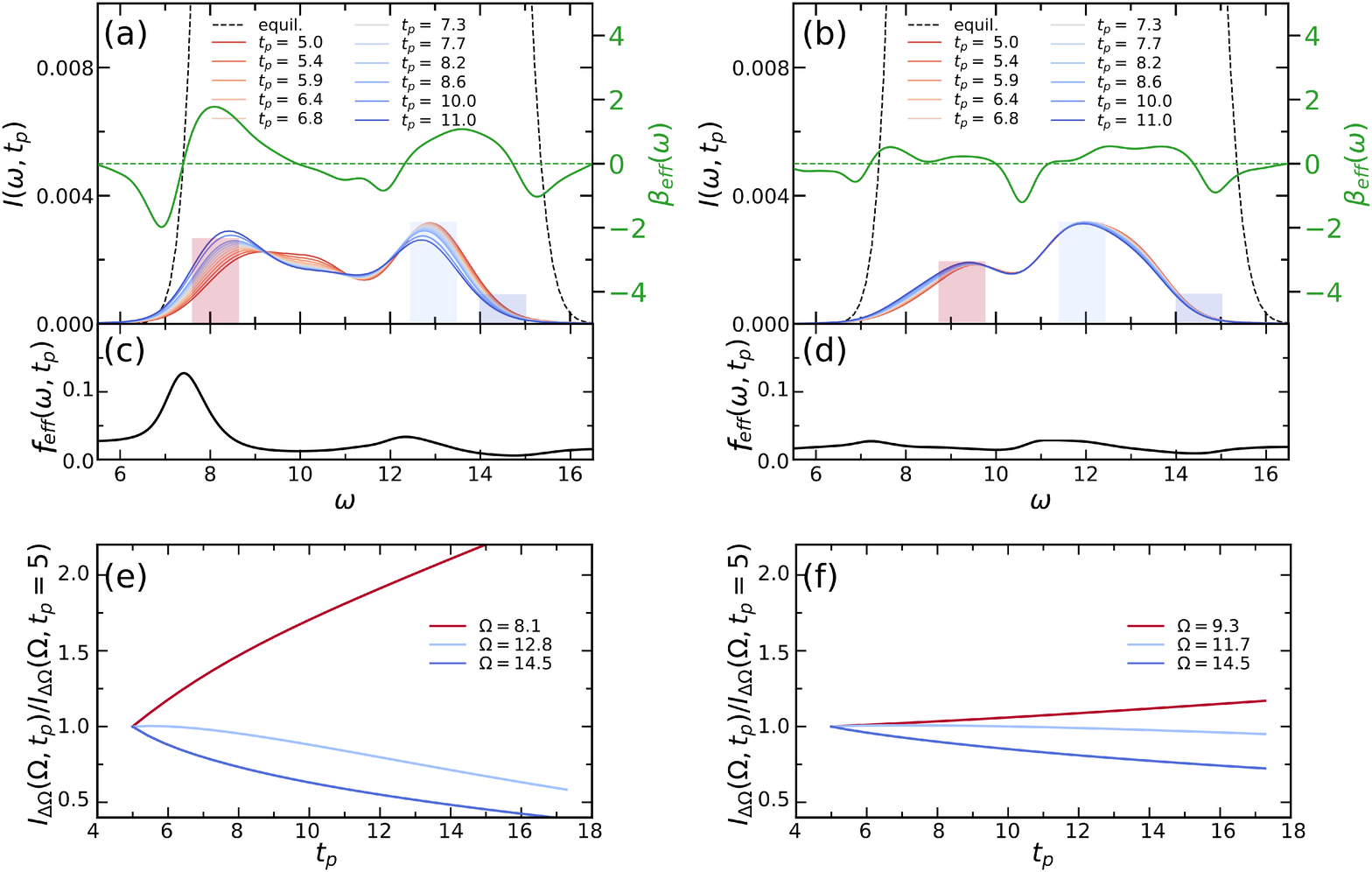}
\caption{(a)-(b) 
Occupation in the upper Hubbard band for a tilted square lattice with $t_h^\prime=0$ (left) and for a triangular lattice with $t_h^\prime=t_h$ (right) after photo-doping (colored solid lines) obtained using the DCA approximation. The black dashed line corresponds to the spectral function calculated from the retarded component of the Green's function in equilibrium, whereas the green solid line corresponds to the effective inverse temperature $\beta_\text{eff}$ calculated at $t_p\approx 11$. (c)-(d) Effective distribution function calculated at $t_p\approx 11$. (e)-(f) Time-dependent photo-emission spectrum integrated in the frequency window $[\Omega-\Delta\Omega,\Omega+\Delta\Omega]$ (shown by shaded areas in (a)-(b) with $\Delta\Omega=0.5$) and normalized to the value at time $t_p=5$ for a tilted square lattice and a triangular lattice, respectively.
}
\label{fig:NonEqAwlesBetaeff}
\end{figure*}

\subsubsection{Double occupancy}
\label{sSec:NonEqDocc}

First, we compare the time-dependent double occupancy during and after the photo-doping pulse in the tilted square lattice ($t_h^\prime=0$) and the triangular lattice ($t_h^\prime=t_h$). The nonequilibrium DCA results are shown by the green and red solid lines in Fig.~\ref{fig:NonEqdocc} (a), respectively. The excitation intensity in these calculations has been adjusted in such a way that the increase in the long-lived doublons, measured at the longest accessible time, is approximately the same in both models, $D(t=18.7)-D_\mathrm{eq} \approx 0.01$. The corresponding pulse amplitudes are $\Delta t_h=0.4$ and $\approx0.47$, respectively. 

Because of the large Mott gap, the lifetime of the photo-doped doublons and holons is much longer than the maximum simulation time,\cite{eckstein2010} so that their density is almost conserved at $t\gtrsim 10$. The photo-excitation of the system is rather short and lasts up to $t\approx 1.8$ (end of the pulse). However, in $D(t)-D_\text{eq}$ we observe strong photo-induced oscillations with a frequency $\approx U$ (see inset in Fig.~\ref{fig:NonEqdocc}(a)) superimposed on an almost step-function-like increase. 
These oscillations originate from hoppings of electrons to nearest-neighbors and back, similar to virtual fluctuations. With increasing frustration (increasing value of $t_h^\prime$), the amplitude of these fluctuations is reduced, since the reduced antiferromagnetic nearest-neighbor correlations enhance the Pauli blocking. The latter effect also explains why a larger pulse strengths is needed in the frustrated lattice to produce the same doublon density as in the square lattice. 

In contrast, the single-site DMFT result in Fig.~\ref{fig:NonEqdocc}(b) shows a very rapid supression of these oscillations after the end of the pump pulse, and an almost identical dynamics for the square and triangular lattice geometry. Hence, one can conclude that geometric frustration, via its effect on the short-range spin correlations and the Pauli blocking, has a significant impact on the doublon dynamics after a short pulse excitation, and in particular leads to a strong damping of the $\omega\approx U$ oscillations.

\subsubsection{Short-range spin correlations}

While the density of doublons and holons is essentially conserved after the pulse, the spin-spin correlations decrease monotonically with increasing time (see Fig.~\ref{fig:noneqSzSz}), apart from small oscillations associated with the previously mentioned nearest-neighbor hopping processes. To compare the dynamics in the models with $t_h^\prime=0$ and $t_h^\prime=t_h$, we normalize the values of $\Snn(t)$ by the equilibrium value, which makes it clear that the decay rate is approximately independent of the degree of frustration. This is consistent with the picture of spin disorder resulting from the hopping of doublons and holons in the initially correlated spin background, assuming that the hopping rate is approximately the same for both values of $t_h^\prime$.   

Qualitatively similar results were previously obtained and discussed in Ref.~\onlinecite{eckstein2016} for the case of a square lattice. The decay of the spin-spin correlation function implies a transfer of (kinetic) energy to the spin sector, which is directly detectable in the time evolution of the doublon distribution function as shown in the following. 

\begin{figure*}[ht]
\centering
	\includegraphics[width=.8\textwidth, draft=false]{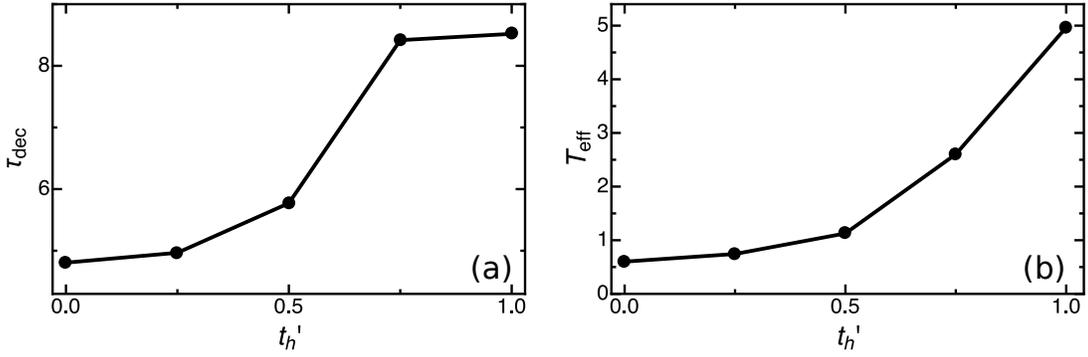}
\caption{
(a) Relaxation time (measured in units of $t_h$) extracted from the PES spectral weight at the upper edge of the UHB as a function of the hopping parameter $t_h^\prime$. 
(b) Effective temperature measured at the lower edge of the UHB at time $t\approx8$.
}
\label{fig:Teff}
\end{figure*}

\subsubsection{Nonequilibrium spectral functions}

For a better understanding of the relaxation dynamics of the photo-excited charge carriers and its dependence on lattice frustration, we simulate the nonequilibrium spectral function (see Eq.~\eqref{eq:spectrum}) in the energy range of the UHB. The results are shown in Fig.~\ref{fig:NonEqAwlesBetaeff} (a)-(b), for several times after the excitation (which lasts up to $t \approx 1.8$). The left panel corresponds to the tilted square lattice, whereas the right panel shows the result for the triangular lattice with $t_h^\prime=t_h$. The frequency of the excitation pulse is chosen in such a way that the short pulse generates a broad and rather uniform occupation in the UHB with a peak closer to its upper edge. After the pulse the corresponding spectral weight is redistributed within the UHB as shown in Fig.~\ref{fig:NonEqAwlesBetaeff} (a)-(b) by the colored solid lines. To analyze more quantitatively the redistribution of the spectral weight within the Hubbard band, we integrate the nonequilibrium spectral function in 
frequency windows of width $2\Delta\Omega$, centered at different $\Omega$:
\beq{
\label{eq:IpesInt}
	I_{\Delta\Omega}(\Omega,t_p)=\int_{\Omega-\Delta\Omega}^{\Omega+\Delta\Omega}I(\omega,t_p)d\omega.
}

The results are plotted in Fig.~\ref{fig:NonEqAwlesBetaeff}(c)-(d) for the tilted square lattice and triangular lattice, respectively. 
Here, the curves are normalized to the value at probe time $t_p=5$ and we use $\Delta\Omega=0.5$. As one can see, in both models the spectral weight is shifting
towards lower frequencies,
 i.e., the photo-doped doublons lose kinetic energy. 
However, the occupation transfer in the tilted square lattice is much more pronounced than in the triangular case. 
 For example, the relative decrease of the spectral weight at the upper band edge ($\Omega=14.5$) between $t_p=5$ and $t_p=14$ is about $50\%$ in the tilted square lattice, while it is only $\sim 20\%$ in the triangular lattice. Even more remarkable is the difference between the two models when we consider the relative increase in spectral weight near the lower band edge where the increase is $> 100\%$ in the square lattice as compared to $\sim 10\%$ in the triangular lattice.
 In contrast, in DMFT simulations (see Fig.~\ref{fig:NonEqAwlesBetaeffDMFT} in Appendix~\ref{sec_appendixA}), where short-range correlations do not play any significant role, the distribution of the spectral weight in the UHB hardly changes after the photo excitation for both the square and triangular lattices.~\cite{eckstein2011}

For a more quantitative analysis of the relaxation dynamics 
 in systems with different degrees of frustration, we extracted the relaxation time from the time-dependent spectral weight at the upper edge of the UHB. The corresponding data are fitted in the time interval $t\in [6, 10]$ by a single exponential function of the form:
\beq{
	I_{\Delta\Omega}(t)=A+B\exp(-t/\tau_\mathrm{dec})
}
where $A$ and $B$ are fitting parameter and $\tau_\mathrm{dec}$ is the relaxation time. The results are plotted in Fig.~\ref{fig:Teff}(a) as a function of the hopping parameter $t_h^\prime$ (degree of geometric frustration). As one can see, increasing frustration leads to a monotonous increase of the relaxation time, which almost doubles for $t_h^\prime=t_h$ compared to the case with $t_h^\prime=0$. This illustrates the  reduced coupling between the charge and spin degrees of freedom in the frustrated system.

\subsubsection{Effective temperature}

To characterize the non-thermal steady state after the photo-excitation, we calculate the effective temperature from the fluctuation-dissipation theorem~\cite{hermann2016,golez2016}
\beq{
	h(\omega, t_p)=\log[-\mathrm{Im}G^R(\omega, t_p)/\mathrm{Im}G^<(\omega, t_p)-1]
}
with $G^R$ and $G^<$ denoting the retarded and lesser components of the nonequilibrium Green's function. The Fourier time window for the calculation of $G^R(\omega, t_p)$ and $G^<(\omega, t_p)$ (analagous to Eq.~\eqref{eq:A} for fixed $t_p$) is set to 10. The slope of $h(\omega, t_p)$ defines a 
frequency-dependent 
effective inverse temperature $\beta_\mathrm{eff}=1/T_\mathrm{eff}$. The results measured at $t\approx 11$ for the tilted square and triangular lattice are shown by the solid green line in Fig.~\ref{fig:NonEqAwlesBetaeff}(a)-(b), respectively. As one can see, $\beta_\mathrm{eff}(\omega)$ shows a strong $\omega$-dependence and even sign changes, especially in the region of the ``pseudo-gap" in $A(\omega)$ (which may be overestimated in DCA due to the piece-wise constant self-energy in momentum space). It is clear though that the effective inverse temperature in the triangular lattice is substantially lower ($T_\text{eff}$ is higher) than in the tilted square lattice, consistent with the broader occupation function, the much weaker accumulation of spectral weight at the lower band edge, and the correspondingly flatter effective distribution function $f_\text{eff}(\omega,t_p)=\text{Im}G^<(\omega,t_p)/\text{Im}G^R(\omega,t_p)$ shown in the panels (c)-(d) of Fig.~\ref{fig:NonEqAwlesBetaeff} for $t_p=11$. For a quantitative comparison, we 
focus on the lowest energy peak in the nonequilibrium spectral function and extract $\beta_\mathrm{eff}$ at the corresponding energy. The resulting $T_\text{eff}=1/\beta_\text{eff}$ is plotted in Fig.~\ref{fig:Teff} for different values of the hopping parameter $t_h^\prime$ (degree of geometrical frustration in the system). As one can see, the effective temperature $T_\mathrm{eff}$ of the photo-doped doublons increases systematically as $t_h^\prime$ is increased. These results demonstrate that the cooling effect associated with energy transfer to the spin background is much reduced in a geometrically frustrated system. This is confirmed by the DMFT simulations (see Fig.~\ref{fig:NonEqAwlesBetaeffDMFT} (a)-(b) in Appendix~\ref{sec_appendixA}), where the absence of spin cooling leads to an effective inverse temperature $\beta_\mathrm{eff}$ close to zero or even negative, independent of the degree of lattice frustration.

\section{Discussion and conclusions}
\label{sec:Summary}

In this work we investigated the effects of geometrical frustration on the relaxation dynamics of photo-doped carriers in the paramagnetic Mott phase of half-filled two-dimensional Hubbard models. We simulated the real-time dynamics of these systems by using the nonequilibrium extension of the dynamical cluster approximation 
in combination with an NCA impurity solver. The use of a $2\times2$ cluster with variable diagonal hoppings allowed us to interpolate between the unfrustrated square lattice and fully frustrated triangular lattice limits, and thus to analyze the effects of geometrical frustration on the short range spin-correlations and time-resolved photoemission spectra.

We showed that geometric frustration manifests itself in different nonequilibrium probes. On the one hand, the increased Pauli blocking in a system with suppressed antiferromagnetic spin correlations leads to suppressed oscillations in the double occupation after a photo-doping pulse, and a reduced energy absorption from the pulse. After doublons have been created with high kinetic energy 
(with a population centered in the upper half of the UHB) 
the geometric frustration has a significant effect on the dissipation of kinetic energy to the spin background and hence the time evolution of the energy distribution function and effective temperature of the doublons. These can serve as fingerprints of the geometric frustration in the nonequilibrium studies.

The spin-charge coupling results in a relaxation of the photo-doped carriers to lower energies and a simultaneous reduction in the nearest-neighbor spin correlations.  Consistent with the results of Ref.~\onlinecite{eckstein2016}, we found that this spin cooling of the photo-doped carriers is very effective in the square lattice case. By systematically varying the hopping parameters we further showed that geometric frustration has a significant effect on the dynamics, since it leads to a reduced energy dissipation at a given hopping rate. As a result, the triangular lattice system exhibits an intra-Hubbard band thermalization of the doublons at a much higher effective temperature than the square lattice system. This effect is not captured in single-site DMFT simulations, where the short-range spin correlations are neglected.

Our results show that nonlocal effects, such as geometrical frustration, not only affect the low energy properties of correlated electron systems, but also play an important role in highly excited nonequilibrium states. This highlights the need for developing computational methods for nonequilibrium systems which go beyond the local DMFT approximation. \\

\acknowledgements

This work was supported by ERC Consolidator Grant No.~724103 (NB, PW), ERC starting grant No.~716648 (ME), and Swiss National Science Foundation Grant No.~200021-165539 (PW). The calculations have been performed on the Beo04 cluster at the University of Fribourg. The Flatiron Institute is a division of the Simons Foundation. We thank M.~Imada for helpful discussions.

\appendix

\section{DMFT}
\label{sec_appendixA}

\begin{figure*}[ht]
\centering
\includegraphics[width=0.9\textwidth, draft=false]{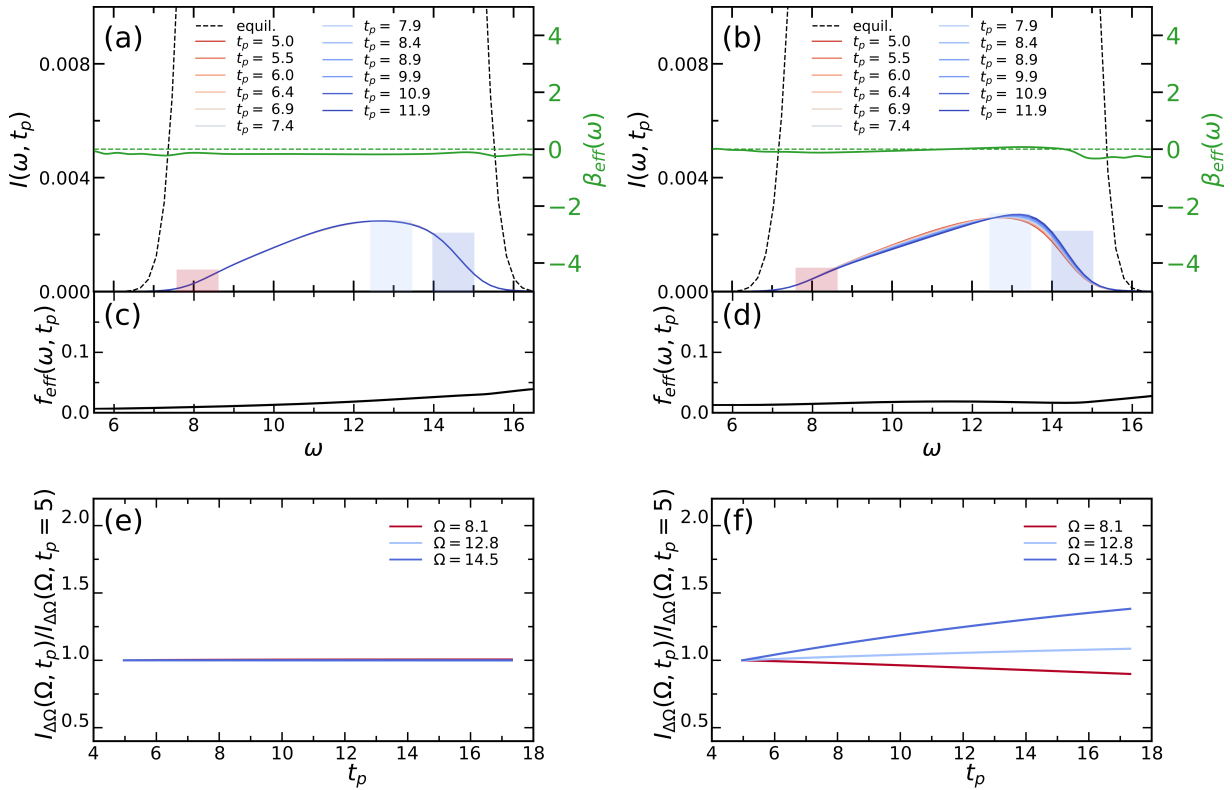}
\caption{(a)-(b) Occupation in the upper Hubbard band for a square lattice (left) and for a triangular lattice with $t_h^\prime=t_h$ (right) after photo-doping (coloured solid lines) obtained using the DMFT approximation. The black dashed line corresponds to the spectral function calculated from the retarded component of the Green's function in equilibrium, whereas the green solid line corresponds to the effective inverse temperature $\beta_\text{eff}$ calculated at $t_p\approx 11$. (c)-(d) Effective distribution function calculated at $t_p\approx 11$. (e)-(f) Time-dependent photoemission spectrum integrated in the frequency window $[\Omega-\Delta\Omega,\Omega+\Delta\Omega]$ (shown by shaded areas in (a)-(b) with $\Delta\Omega=0.5$) and normalized relative to the value at $t_p=5$ for a square lattice and a triangular lattice, respectively. The excitation protocols are the same as in Fig.~\ref{fig:NonEqdocc} (b).
}
\label{fig:NonEqAwlesBetaeffDMFT}
\end{figure*}

For the purpose of comparison with the DCA results in Fig.~\ref{fig:NonEqAwlesBetaeff}, we show in Fig.~\ref{fig:NonEqAwlesBetaeffDMFT} the corresponding nonequilibrium photoemission spectra  (Eq.~\eqref{eq:spectrum})
in the energy range of the UHB obtained using the single-site DMFT approximation. The left panels correspond to the case of a square lattice, whereas the right panels show the results for a triangular lattice ($t_h^\prime=t_h$). The photo-excitation protocol is the same as discussed in Sec.~\ref{sSec:NonEqDocc} in connection with Fig.~\ref{fig:NonEqdocc}(b). 

As one can see in Fig.~\ref{fig:NonEqAwlesBetaeffDMFT} (a)-(b), the photo-induced spectral weight in the UHB does not significantly change with time for both lattice geometries. This can be explained by a very fast relaxation towards a non-thermal steady state due to the lack of short-range spin correlations. The calculated inverse effective temperature $\beta_\mathrm{eff}$ in this state at $t\approx 11$ (green line) shows in both cases almost frequency independent values close to zero (similar to the effective distribution function shown in the panels (c)-(d)), 
which is also consistent with the absence of spin cooling. 

To get more quantitative insights into the dynamics of the spectral weight within the UHB, we calculate the integrated spectral function $I_{\Delta\Omega}(\Omega,t_p)$ according to Eq.~\eqref{eq:IpesInt}. The normalized results with respect to $ I_{\Delta\Omega}(\Omega,t_p=5)$ are plotted in Fig.~\ref{fig:NonEqAwlesBetaeffDMFT} (e)-(f). 
Whereas the system in the square lattice case shows no changes in $I_{\Delta\Omega}(\Omega,t_p)$ with probing time $t_p$, the spectral weight for the triangular lattice geometry is shifting from the lower edge of the UHB to its upper edge. We note, however, that 
this result is obtained from the division of two very small numbers. 
In absolute values the time-dependent changes are tiny.

\bibliography{Polarons,tdmft,Books}

\end{document}